\documentclass{jpp}

\newcommand{\bB}{\boldsymbol{B}}
\newcommand{\bv}{\boldsymbol{v}}
\newcommand{\bA}{\boldsymbol{A}}

\newcommand{\bJ}{\boldsymbol{J}}
\newcommand{\nn}{\nonumber}

\newcommand{\nb}{\bnabla}

       \newcommand{\Cc}{ {\mathcal{C}} }

       \newcommand{\Fc}{ {\mathcal{F}} }
       \newcommand{\Gc}{ {\mathcal{G}} }
       \newcommand{\Hc}{ {\mathcal{H}} }

       \newcommand{\Lc}{ {\mathcal{L}} }
       \newcommand{\Mc}{ {\mathcal{M}} }
       \newcommand{\Nc}{ {\mathcal{N}} }

       \newcommand{\Rc}{ {\mathcal{R}} }

       \newcommand{\Wc}{ {\mathcal{W}} }

       \newcommand{\R}{ {\mathbb{R}} }
       
       \newcommand{\bh}{ {\boldsymbol{h}} }

\usepackage{graphicx}
\usepackage{epstopdf, epsfig}
\usepackage{hyperref}
\usepackage{natbib}

\usepackage{amsfonts}
\usepackage[T1]{fontenc}
\usepackage{mathtools}
\usepackage{amsfonts}
\usepackage{dcolumn}
\usepackage{bm}
\usepackage[export]{adjustbox}
\usepackage{wrapfig}
\usepackage{lipsum}
\usepackage[titletoc,toc,title]{appendix}

\shorttitle{Helically symmetric extended MHD}
\shortauthor{D. A. Kaltsas, G. N. Throumoulopoulos and P. J. Morrison}

\title{Helically symmetric extended magnetohydrodynamics: Hamiltonian formulation and equilibrium variational principles}

\author{D. A. Kaltsas\aff{1}, G. N. Throumoulopoulos\aff{1}  \corresp{\email{gthroum@cc.uoi.gr}}
 \and P. J. Morrison\aff{2}}

\affiliation{\aff{1}Department of Physics, University of Ioannina,\\ GR 451 10 Ioannina, Greece
\aff{2}Department of Physics and Institute for Fusion Studies,\\ University of Texas, Austin, Texas 78712, USA }

\begin{document}

\maketitle

\begin{abstract}
Hamiltonian extended magnetohydrodynamics (XMHD) is restricted to respect helical symmetry by reducing the Poisson bracket for 3D dynamics to a helically symmetric one, as an extension of the previous study for translationally symmetric XMHD (D.A. Kaltsas et al, Phys. Plasmas 24, 092504 (2017)). Four families of Casimir invariants are obtained directly from the symmetric Poisson bracket and they are used to construct Energy-Casimir variational principles for deriving generalized XMHD equilibrium equations with arbitrary macroscopic flows. The system is then cast into  the form of Grad-Shafranov-Bernoulli equilibrium equations. The axisymmetric and the translationally symmetric formulations can be retrieved  as geometric reductions of the helical symmetric one. As special cases, the derivation of the corresponding equilibrium equations for incompressible plasmas is discussed and the helically symmetric equilibrium equations for the Hall MHD system are obtained upon neglecting electron inertia. An example of an incompressible double-Beltrami equilibrium is presented in connection with a magnetic configuration having non-planar helical magnetic axis. 
\end{abstract}

\section{Introduction}
\label{sec_I}
Extended magnetohydrodynamics (XMHD) is perhaps the simplest consistent, in terms of energy conservation~\citep{Kimura2014}, fluid plasma model containing both Hall drift and electron inertial effects. It is obtained by reduction of the standard two-fluid plasma model,  when the quasineutrality  assumption  is imposed  and expansion in the smallness of the electron-ion mass ratio is performed ~\citep{Lust1959,Kimura2014}, although the latter expansion need not be done (see Sec.~VI of  \citet{Kawazura2017}). The Hamiltonian structure of this model was first identified in \citet{Abdelhamid2015}  for its barotropic version and corroborated in  \citet{Lingam2015},  where transformations  to the Hamiltonian structures  of Hall MHD (HMHD) (e.g.\ \citet{Lighthill1960}), Inertial MHD (IMHD)  \citep{Kimura2014,Lingam2015a} and the ordinary ideal  MHD model were identified.  The Hamiltonian structure of XMHD served as  the starting points for two subsequent papers  that dealt with applications of its translationally symmetric counterpart to magnetic reconnection \citep{Grasso2017} and equilibria \citep{Kaltsas2017}. In the former publication the incompressible case with homogeneous mass density was considered,  while in the latter the analysis concerned the compressible barotropic version of the model.   

Here we present the Hamiltonian formulation of the barotropic XMHD model in the presence of  continuous helical symmetry,   an extension of our previous work \citep{Kaltsas2017} that was concerned with translationally symmetric plasmas.  Helical symmetry is a general case  that includes both axial and translational symmetry. Therefore the results obtained within the context of a helically symmetric formulation can be directly applied to the sub-cases of axial and translational symmetry. This provides a unified framework for the study of equilibrium and stability of symmetric configurations,  which is important because  purely or nearly helical structures are very common in plasma systems. For example, 3D equilibrium states with internal helical structures with toroidicity, e.g.,  helical cores, have been observed experimentally \citep{Weller1987,Pecquet1997} and simulated \citep{Cooper2010,Cooper2011} in Tokamaks and RFPs (e.g.\  \citet{Lorenzini2009,Bergerson2011,Puiatti2009,Terranova2010}).  Another example of helical structures that emerge from  plasma instabilities, such as  the resistive or collisionless tearing modes, or as a result of externally  imposed  symmetry-breaking  perturbations are  magnetic islands \citep{Waelbroeck2009}. In addition the helix may serve as a rough approximation of helical non-axisymmetric devices \citep{Uo1961} and can be useful to investigate some features of stellarators \citep{Spitzer1958,Helander2012}, the second major class of magnetic confinement devices alongside the Tokamak, in the large aspect-ratio limit. Also helical magnetic structures are common in astrophysics, e.g.,  in astrophysical jets \citep{Pino2005,Pudritz2012}. Therefore it is of interest to derive a joint tool for two-fluid equilibrium and stability studies of systems with helical symmetry, with the understanding that for   most cases of laboratory applications, helical symmetry is an idealized approximation.

As in our previous work \citep{Kaltsas2017}, we use the Energy-Casimir (EC) variational principle to obtain equilibrium conditions. However it is known that the EC principle can be extended for the study of linear, and nonlinear stability \citep{Holm1985,Morrison1998} by investigating the positiveness of the second variation of the EC functional, an idea that dates to the early plasma literature \citep{Kruskal1958}. Many works that employ such principles for the derivation of equilibrium conditions and sufficient MHD stability criteria, arising as consequences of the noncanonical Hamiltonian structure of ideal MHD \citep{Morrison1980}, have been published over the last decades for several geometric configurations \citep{Holm1985,Holm1987,Almaguer1988,Morrison2013,Moawad2013,Andreussi2013,Andreussi2016,Moawad2017}. In \citet{Andreussi2012,Andreussi2013} Energy-Casimir equilibrium and stability principles were used in the case of helically symmetric formulation. A similar equilibrium variational principle was applied in the case of XMHD  \citep{Kaltsas2017} for plasmas with translational symmetry. Therefore the use of such principles in the case of helically symmetric XMHD seems a natural generalization of the previous studies. To accomplish this task we first derive the Poisson bracket of the helically symmetric barotropic XMHD and its corresponding families of Casimir invariants. Those invariants, along with the symmetric version of the Hamiltonian function are used in an Energy-Casimir variational principle in order to obtain the equilibrium equations for helical plasmas described by   XMHD. To our knowledge this is the first time that equilibrium equations containing two-fluid physics are derived for helical configurations, especially exploiting Hamiltonian techniques. 

The present study is organized as follows: in Sec.~\ref{sec_II} we present briefly the Hamiltonian field theory of barotropic XMHD.  In Sec.~\ref{sec_III} we introduce the requisite  description of the helical coordinate and representations of the helically symmetric magnetic  and  velocity fields.  Then, the XMHD Poisson bracket is reduced to its helically symmetric counterpart. In Sec.~\ref{sec_IV} the Casimir invariants of the symmetric bracket are obtained and their MHD limit is considered. Also we establish the symmetric EC variational principle, from which we  derive generalized equilibrium equations for helical systems. Special cases of equilibria such as the Hall MHD equilibria, are discussed in detail in Sec.~\ref{sec_V}. We conclude with Sec.~\ref{sec_VI}, where we discuss the results of our study. 
\section{Barotropic XMHD}
\label{sec_II}
\subsection{Evolution equations}
The barotropic XMHD equations, presented in a series of recent articles \citep{Kimura2014,Abdelhamid2015,Lingam2015,Grasso2017,Kaltsas2017}, in Alfv\'en units, are given by:
\begin{eqnarray}
&&\partial_{t}\rho=-\bnabla\bcdot\left(\rho \bv\right)\,,
\label{ConDen} \\
&&\partial_t \bv=\bv\times(\bnabla\times\bv)-\bnabla v^2/2-\rho^{-1}\nb p+\rho^{-1}\bJ\times\bB^*
  -d_e^2\bnabla\left(\frac{|\bJ|^2}{2\rho^2}\right)\,, \label{mom_eq}\\
&&\partial_t\boldsymbol{B}^*=\bnabla\times\left(\bv\times\bB^*\right)-d_i\bnabla\times\left(\rho^{-1}\bJ\times \bB^*\right) +d_e^2\bnabla\times\left[\rho^{-1}\bJ\times\left(\bnabla\times\bv\right)\right] \,,
\label{starB}
\end{eqnarray}
where
\begin{eqnarray}
\bJ=\bnabla\times\bB\,,\qquad \bB^*&=&\bB+d_e^2\bnabla\times\left(\frac{\nb\times\bB}{\rho}\right)\,. \label{B*}
\end{eqnarray}
The parameters $d_i$ and $d_e$ are the normalized ion and electron skin depths respectively, $p=p(\rho)$ is the total pressure and $\rho$, $\bv$ and $\bB$ represent the mass density, the velocity and the magnetic field, respectively.
\subsection{Hamiltonian formulation}
It has been recognized that the equations \eqref{ConDen}-\eqref{starB} possess a noncanonical Hamiltonian structure, i.e. the dynamics can be described by a set of generalized Hamiltonian equations  \citep{Morrison1982,Morrison1998}  
\begin{equation}
\partial_t\eta=\{\eta,\Hc\}\,, \label{gen_ham_eq}
\end{equation}
where $\eta=(\rho,\bv,\bB^*)$ are noncanonical dynamical variables (not consisting of canonically conjugate pairs), $\Hc[\rho,\bv,\bB^*]$ is a real valued Hamiltonian functional,  and $\{F,G\}$ is a Poisson bracket acting on functionals of the variables $\eta$, which is bilinear, antisymmetric,  and satisfies the Jacobi identity. The appropriate Hamiltonian for our system is the following: 
\begin{eqnarray}
\Hc=\int_D d^3x\,\left[\rho \frac{v^2}{2}+\rho U(\rho)+\frac{\bB\bcdot\bB^*}{2}\right]\,,
 \label{hamiltonian}
\end{eqnarray} 
where $D\subseteq \R^3$ and $U$ is the internal energy function ($p=\rho^2dU/d\rho$), while the corresponding noncanonical  Poisson bracket is
\begin{eqnarray}
\left\{F,G\right\}&=&\int_D  d^3x\,\big\{G_\rho\bnabla\bcdot F_{\bv} - F_\rho\bnabla\bcdot G_{\bv}
+\rho^{-1}\left(\bnabla\times\bv\right)\bcdot\left(F_{\bv} \times G_{\bv}\right) \nn \\
&& \hspace{1.2cm}+ \rho^{-1}\bB^*\bcdot\left[F_{\bv}\times\left( \bnabla\times G_{\bB^*}\right)-G_{\bv}\times \left(\bnabla\times F_{\bB^*}\right)\right] \nn \\
&&\hspace{1.2cm} - d_i \rho^{-1}\bB^*\bcdot\left[\left(\bnabla\times F_{\bB^*}\right)\times \left(\bnabla\times G_{\bB^*}\right)\right]\nn \\
&& \hspace{1.2cm} + d_e^2\rho^{-1}\left(\bnabla\times \bv\right)\bcdot \left[\left(\bnabla\times F_{\bB^*}\right)\times \left(\bnabla\times G_{\bB^*}\right)\right]\big\}\,, \label{poisson}
\end{eqnarray}
%
where $F_{\boldsymbol{z}}:={\delta F}/{\delta \boldsymbol{z}}$ denotes the functional derivative of $F$ with respect to the dynamical variable $\boldsymbol{z}$, defined by $\delta F[\boldsymbol{z},\delta\boldsymbol{z}]=\int_D d^3x\,\delta\boldsymbol{z}\bcdot(\delta F/\delta\boldsymbol{z}) \,$. For the computation of the functional derivatives of the field variables we make use of $\delta z_i(\boldsymbol{x}')/\delta z_j(\boldsymbol{x})=\delta_{ij}\delta(\boldsymbol{x}'-\boldsymbol{x})$.

For noncanonical (degenerate) Poisson brackets, such as the bracket \eqref{poisson}, there exist functionals $\Cc[\eta]$ that commute with every arbitrary functional $F[\eta]$
\begin{equation}
\{F,\Cc\}=0\,, \quad \forall F\,.
\end{equation} 
These functionals $\Cc$ are called Casimir invariants and obviously they do not change the dynamics if $\Hc\rightarrow\mathfrak{F}=\Hc-\sum_i \Cc_i$, that is
\begin{equation}
\partial_t\eta=\{\eta,\mathfrak{F}\}\,,
\end{equation}
describes the same dynamics as \eqref{gen_ham_eq}. 

Equilibrium solutions satisfy $\{\eta,\mathfrak{F}\}=0$, which is true if the first variation of the generalized Hamiltonian functional $\mathfrak{F}$ vanishes at the equilibrium point, i.e., 
\begin{equation}
\delta \mathfrak{F}=
\delta\left(\Hc-\sum_i \Cc_i\right)=0 \,,
\label{en_cas}
\end{equation}
is a sufficient but not necessary condition for equilibria \citep{Morrison1998,Yoshida2014}. To obtain stability criteria one may take the second variation of the EC functional. It is known that if the second variation $\delta^2\mathfrak{F}$ at the equilibrium point is positive definite, then it provides a norm which is conserved by the linear dynamics, so the equilibrium is linearly stable \citep{Kruskal1958,Holm1985,Morrison1998}. 

The aim of the following sections is to derive the Casimir invariants of the helically symmetric XMHD and then to find the corresponding equilibrium equations via the condition \eqref{en_cas}. For the general 3D version of the model described by means of \eqref{hamiltonian} and \eqref{poisson}, the Casimir invariants are 
%
\begin{eqnarray}
\Cc_1&=&\int_D d^3x\, \rho \,,\\
\Cc_{2,3}&=&\int_D d^3x\, \left(\bA^*+\gamma_{\pm}\bv\right)\bcdot \left(\bB^*+\gamma_{\pm}\bnabla\times\bv\right)\,, \label{cas_xmhd}
\end{eqnarray}
%
with $\bB^*=\bnabla\times\bA^*$ and $\gamma_{\pm}$ being the two roots of the quadratic equation $\gamma^2-d_i\gamma-d_e^2=0$, i.e. $\gamma_\pm= \left(d_i\pm\sqrt{d_i^2+4d_e^2}\right)/2$.
\section{Helically symmetric formulation}
\label{sec_III}
As mentioned above, the helically symmetric formulation includes both the translationally symmetric and axisymmetric cases, while being the most generic case for which a poloidal representation of the magnetic field is possible, i.e. a global description in terms of a component parallel to the symmetry direction and a flux function describing the field that lies on a plane perpendicular to this direction (poloidal plane), a representation which provides well defined magnetic surfaces.
In a series of papers this symmetry was employed for deriving  equilibrium equations of the Grad-Shafranov type, i.e. PDEs with solutions being poloidal magnetic flux functions, \citep{Johnson1958,Tsinganos1982,Bogoyavlenskij2000,Throumoulopoulos1999,Andreussi2012,Evangelias2018} in the context of standard MHD theory. Particularly in \citet{Andreussi2012} the equilibrium Grad-Shafranov or JFKO \citep{Johnson1958,Bogoyavlenskij2000} equation was derived using a Hamiltonian variational principle. The same approach is adopted also for our derivation, however, for the more complicated XMHD theory. 
\subsection{Helical symmetry and Poisson bracket reduction}
The helical symmetry can be imposed by assuming that in a cylindrical coordinate system $(r,\phi,z)$ all equations of motion depend spatially on $r$ and the helical coordinate $u=\ell \phi+n z$, where $\ell=\sin (a)$ and $n=-\cos(a)$ with $a$ being the helical angle. For $a=0$ we obtain the axisymmetric case and for $a=\pi/2$ the translationally symmetric case. The contravariant unit vector in the direction of the $u$ coordinate is $\boldsymbol{e}_u=\nb u/|\nb u|=\ell k \boldsymbol{e}_\phi+ nkr\boldsymbol{e}_z$,  where $k$ is 
\begin{equation}
k:=\frac{1}{\sqrt{\ell^2+n^2r^2}}\,.
\end{equation}
The tangent to the direction of the helix $r = const.$ $u = const.$ is given by $\boldsymbol{e}_h=\boldsymbol{e}_r\times \boldsymbol{e}_u$ and one can prove that the following relations hold:
%
\begin{equation}
\nb\bcdot \bh=0\,,\quad \nb\times\bh=-2n\ell k^2 \bh\,, \label{div_curl_h}
\end{equation}
%
where $\bh=k\boldsymbol{e}_h$, hence $\bh\bcdot\bh=k^2$. Helical symmetry means that $\bh\bcdot\bnabla f=0$ where $f$ is arbitrary scalar function. The relations \eqref{div_curl_h} give us the opportunity to introduce the so-called poloidal representation for the divergence-free magnetic field and also a poloidal representation for the velocity field, adding though a potential field contribution accounting for the compressibility of the velocity field, i.e., 
%
\begin{eqnarray}
&&\bB^*=k^{-1}B_h^*(r,u,t)\bh+\bnabla\psi^*(r,u,t)\times\bh\,, \label{hel_sym_magn_field}\\
&&\bv=k^{-1}v_h(r,u,t)\bh+\bnabla\chi(r,u,t) \times \bh+\bnabla \Upsilon(r,u,t) \,.
\label{hel_sym_vel_field}
\end{eqnarray}
%
For incompressible flows $\Upsilon$ is harmonic or constant.  
In view of \eqref{div_curl_h},  the divergence and the curl of (\ref{hel_sym_magn_field}) and  \eqref{hel_sym_vel_field} are given by 
%
\begin{eqnarray}
\bnabla\bcdot\bv&=&\Delta\Upsilon\,,\quad \nb\bcdot \bB^*=0\,, \\
\bnabla\times \bv&=&\left[k^{-2}\Lc\chi-2n\ell kv_h\right]\bh+\nb(k^{-1}v_h)\times\bh \,,\label{curl_v}\\
\nb\times \bB^*&=&\left[k^{-2}\Lc\psi^*-2n\ell k B_h^*\right]\bh+\nb(k^{-1}B_h^*)\times\bh \,,
\end{eqnarray}
%
where $\Delta:=\nb^2$ and $\Lc:=-\nb\bcdot(k^2\nb (\bcdot))$ is a linear, self-adjoint differential operator. 
For convenience we define the following quantities:  $w:=\Delta \Upsilon$ or  $\Upsilon=\Delta^{-1}w$ and $\Omega=\Lc \chi$  or $\chi=\Lc^{-1}\Omega$. 

Having introduced the representation of  \eqref{hel_sym_magn_field}--\eqref{hel_sym_vel_field} for the helically symmetric fields, in order to derive the helically symmetric Hamiltonian formulation we need to express the Hamiltonian \eqref{hamiltonian} and the Poisson bracket \eqref{poisson} in terms of the scalar field variables $\eta_{_{HS}}=(\rho,v_h,\chi,\Upsilon, B_h^*,\psi^*)$. This is achieved not only by expressing the fields $\eta_{_{3D}}=(\rho,\bv,\bB^*)$ in terms of the scalar field variables but it requires also the transformation of the functional derivatives from derivatives with respect to $\eta_{_{3D}}$ to functional derivatives with respect to the  scalar fields $\eta_{_{HS}}$.  As in \citet{Andreussi2010,Andreussi2012,Kaltsas2017},  we  accomplish this transformation by employing a chain rule reduction,
%
\begin{eqnarray}
F_\rho=F_\rho\,,\quad F_{\bv}=k^{-1}F_{v_h}\bh+\nb F_\Omega\times \bh-\nb F_w\,, \label{F_v} \\
F_{\bB^*}=k^{-1}F_{B_h^*}\bh-k^{-2}\nb\left(\Delta^{-1}F_{\psi^*}\right)\times\bh\,, \label{F_B} 
\end{eqnarray}
%
where
\begin{equation}
F_w=\Delta^{-1} F_\Upsilon\, , \qquad F_\Omega=\Lc^{-1} F_\chi\, ,
\end{equation}
%
which follow from 
%
\begin{eqnarray}
\int_D d^3x\, F_\chi \delta\chi =\int_D d^3\, x F_\Omega \delta \Omega  \,, \\
\int_D d^3x\,  F_\Upsilon \delta\Upsilon =\int _Dd^3x\,  F_w \delta w  \,,
\end{eqnarray}
%
upon introducing the relations $\delta\Omega=\Lc\delta\chi$, $\delta w=\Delta \delta \Upsilon$ and exploiting the self-adjointness of the operators $\Delta$ and $\Lc$.
Also we observe that in \eqref{poisson} there exist bracket blocks which contain the curl of $F_{\bB^*}$, which is 
%
\begin{equation}
\nb\times F_{\bB^*}=\left(k^{-2}F_{\psi^*}-2n\ell  k F_{B_h^*}\right)\bh+\nb\left(
k^{-1}F_{B_h^*}\right)\times\bh\,. \label{curl_F_B} 
\end{equation}
%
The helically symmetric Poisson bracket occurs by substituting \eqref{hel_sym_magn_field}, \eqref{curl_v}, \eqref{F_v} and \eqref{curl_F_B} into \eqref{poisson} and assuming that any surface-boundary terms which emerge due to integrations by parts, vanish due to appropriate boundary conditions, for example periodic conditions or vanishing field variables $\eta_{_{HS}}$ on $\partial D$, except for the mass density $\rho$ because various terms diverge as $\rho$ approaches to zero, as is evident even from \eqref{poisson}. However, in view of the actual physical situation one can assume that the mass density on the boundary is sufficiently small. The Poisson bracket takes the form
%
\begin{eqnarray}
\{F,G\}^{^{XMHD}}_{_{HS}}&&=\int_Dd^3x\, \Big\{F_\rho\Delta G_w-G_\rho\Delta F_w +\rho^{-1}\left(\Omega-2n\ell k^3v_h\right)\times \nn \\
&&\times\big(\left[F_{\Omega},G_{\Omega}\right]
+k^{-2}[F_w,G_w]+\nb F_w \bcdot\nb G_\Omega- \nb F_\Omega\bcdot \nb G_w
\big) \nn\\
&&+k^{-1}v_h\big([F_\Omega,\rho^{-1}k G_{v_h}]-[G_\Omega,\rho^{-1}k F_{v_h}]
\nn\\
&& +\nb \bcdot (\rho^{-1}kG_{v_h}\nb F_w) -\nb\bcdot (\rho^{-1}kF_{v_h} \nb G_w)  
\big) \nn \\
&&
+\rho^{-1}kB_h^*\big([F_\Omega ,k^{-1}G_{B_h^*}]-[G_\Omega ,k^{-1} F_{B_h^*}] \nn \\
&&+\nb F_w\bcdot\nb \left(k^{-1}G_{B_h^*}\right)  -\nb G_w\bcdot\nb \left(k^{-1}F_{B_h^*}\right) 
\big) \nn \\
&&+\psi^*\big(
[F_{\Omega},\rho^{-1}G_{\psi^*}]-[G_{\Omega},\rho^{-1} F_{\psi^*}]
+[k^{-1}F_{B_h^*},\rho^{-1}kG_{v_h}]  \nn \\
&& -[k^{-1}G_{B_h^*},\rho^{-1}k F_{v_h}]
+\nb\bcdot\left(\rho^{-1}G_{\psi^*}\nb F_w\right)-\nb\bcdot\left(\rho^{-1}F_{\psi^*}\nb G_w\right) \big) \nn \\
&& -2n\ell \psi^*\big([F_\Omega,\rho^{-1}k^3G_{B_h^*}]-[G_\Omega,\rho^{-1}k^3F_{B_h^*}]\nn \\ 
&&+\nb\left(\rho^{-1}k^3G_{B_h^*}\nb F_w\right)-\nb\left(\rho^{-1}k^3F_{B_h^*}\nb G_w\right)\big)\nn\\
&&-d_i \rho^{-1} k B_h^*[k^{-1}F_{B_h^*},k^{-1}G_{B_h^*}]\nn \\
 &&-d_i \psi^* \big([\rho^{-1}F_{\psi^*},k^{-1}G_{B_h^*}]-[\rho^{-1}G_{\psi^*},k^{-1}F_{B_h^*}]
\big) \nn\\
&&+ 2n\ell d_i\psi^*\big([\rho^{-1}k^3F_{B_h^*},k^{-1}G_{B_h^*}]-[\rho^{-1}k^3G_{B_h^*},k^{-1}F_{B_h^*}]\big)
\nn\\
&&+d_e^2 \rho^{-1}\left(\Omega-2n\ell k^3v_h\right) [k^{-1}F_{B_h^*},k^{-1}G_{B_h^*}]
\nn\\
&& +d_e^2k^{-1}v_h\big([\rho^{-1}F_{\psi^*},k^{-1}G_{B_h^*}]-[\rho^{-1}G_{\psi^*},k^{-1}F_{B_h^*}]\big) \nn\\
&&-2n\ell d_e^2 k^{-1} v_h\big([\rho^{-1}k^3F_{B_h^*},k^{-1}G_{B_h^*}]-[\rho^{-1}k^3G_{B_h^*},k^{-1}F_{B_h^*}]\big)
\Big\} \,,
 \label{hs_poisson}
\end{eqnarray}
%
where  $[f,g]:=\left(\nb f\times\nb g\right)\bcdot\bh$ is the helical Jacobi-Poisson bracket.  One may prove that with appropriate boundary conditions, e.g. such those mentioned above, the identity 
%
\begin{equation}
\int_Dd^3x\, [f,g]h=\int_Dd^3x\, [h,f]g=\int_Dd^3x\,[g,h]f\,,
 \label{identity}
\end{equation}
%
holds for arbitrary functions $f,g,h$. These  conditions are necessary to derive the bracket \eqref{hs_poisson}   and also for finding the Casimir determining equations.

 It's not difficult to show that if we set $a=\pi/2$ the bracket \eqref{hs_poisson} reduces to the translationally symmetric XMHD bracket derived in \citet{Kaltsas2017}. The corresponding axisymmetric bracket can be obtained by setting $a=0$. In this case the purely helical terms which contain a coefficient $2n\ell $ vanish and the scale factor $k$ becomes $1/r$. 

To complete the Hamiltonian description of helically symmetric XMHD dynamics we need to express the Hamiltonian \eqref{hamiltonian} in terms of the scalar fields $\eta_{_{HS}}$,   leading to 
\begin{eqnarray}
\Hc=\int_Dd^3x\, &&\bigg\{\frac{\rho}{2}\left(v_h^2+k^2|\bnabla \chi|^2+|\nb\Upsilon|^2\right)\nn \\
&&+\rho \left([\Upsilon,\chi]+U(\rho)\right)
+\frac{B_h^*B_h}{2}+k^2\frac{\nb\psi^*\bcdot\nb\psi}{2}\bigg\}\,.
\label{hs_hamiltonian}
\end{eqnarray}
Also from the definition of the generalized magnetic field $\bB^*$ \eqref{B*} and the helical representation \eqref{hel_sym_magn_field} one may derive the following relations for the generalized variables $B_h^*$ and $\psi^*$:
\begin{eqnarray}
B_h^*&=& (1+4n^2\ell^2d_e^2\rho^{-1}k^4)B_h+d_e^2\big[\rho^{-1}k^{-1}\Lc(k^{-1}B_h)\nn \\
&&\hspace{1.5cm}-2n\ell \rho^{-1}k\Lc\psi-k\nb\rho^{-1}\bcdot\nb(k^{-1}B_h)\big]\,, \label{B_h*}\\
\psi^*&=&\psi+d_e^2\left[\rho^{-1}k^{-2}\Lc\psi-2n\ell \rho^{-1}kB_h\right]\,,
\label{psi*}
\end{eqnarray}
 where $B_h$ is the helical component and $\psi$ the poloidal flux function of the magnetic field $\bB$.
Note that terms containing the parameters $n$ and $\ell$ are purely helical, i.e.,  they vanish in the cases of axial and translational symmetry. Also the last term of \eqref{B_h*} is purely compressible, i.e.,  it vanishes if we consider incompressible plasmas. Another interesting observation is that due to the non-orthogonality of the helical coordinates, there is a poloidal magnetic field contribution in the helical component of the generalized magnetic field $B_h^*$ and helical magnetic contribution $B_h$ in the poloidal flux function $\psi^*$. This mixing makes the subsequent dynamical and equilibrium analyses appear much more involved than in our previous study, however it can be simplified upon observing that 
\begin{eqnarray}
&&\int_Dd^3x\left[B_h^*\delta B_h+\Lc(\psi^*)\delta\psi\right]\nn \\
&&=\int_Dd^3x\left[B_h\delta B_h^*+\Lc(\psi)\delta\psi^*+\frac{d_e^2}{\rho^2}\left(J_h^2+k^2|\nb(k^{-1}B_h)|^2\right)\delta\rho\right]\,,
\end{eqnarray}
where $J_h=k^{-1}\Lc\psi-2n\ell k^2 B_h$ is the helical component of the current density. Therefore the variation of the magnetic part of the Hamiltonian can be written as
\begin{eqnarray}
   \delta\Hc_m&=&\int_D d^3x\left[\frac{1}{2}B_h^*\delta B_h+\frac{1}{2}B_h\delta B_h^*+\frac{1}{2}\Lc(\psi^*)\delta\psi+\frac{1}{2}\Lc(\psi)\delta\psi^*\right]\nn\\
   &=&\int_D d^3x\left[B_h\delta B_h^*+\Lc(\psi)\delta\psi^*+\frac{d_e^2}{2\rho^2}\left(J_h^2+k^2|\nb(k^{-1}B_h)|^2\right)\delta\rho\right]\nn\\
  &=&\int_D d^3x\left[B_h^*\delta B_h+\Lc(\psi^*)\delta\psi-\frac{d_e^2}{2\rho^2}\left(J_h^2+k^2|\nb(k^{-1}B_h)|^2\right)\delta\rho\right]\,,
\end{eqnarray}
leading to the following relations for the functional derivatives of the Hamiltonian:
\begin{eqnarray}
&&\frac{\delta\Hc}{\delta B_h}=B_h^*\,,\quad 
 \frac{\delta\Hc}{\delta \psi}=\Lc\psi^*\,, \label{func_der_1}\\
&& \frac{\delta\Hc}{\delta B_h^*}=B_h\,,\quad \frac{\delta\Hc}{\delta \psi^*}=\Lc\psi\,, \label{func_der_2}\\ 
&&\frac{\delta\Hc}{\delta \rho}\bigg|_{B_h^*,\psi^*}=\frac{v^2}{2}+\left[\rho U(\rho)\right]_\rho+\frac{d_e^2}{2\rho^2}\left(J_h^2+k^2|\nb(k^{-1}B_h)|^2\right)\,, \label{func_der_3}\\
&&\frac{\delta\Hc}{\delta \rho}\bigg|_{B_h,\psi}=\frac{v^2}{2}+\left[\rho U(\rho)\right]_\rho-\frac{d_e^2}{2\rho^2}\left(J_h^2+k^2|\nb(k^{-1}B_h)|^2\right)\,.\label{func_der_4}
\end{eqnarray}
%
In addition,  the functional derivatives with respect to the velocity related variables are given by
%
\begin{eqnarray}
&&\frac{\delta \Hc}{\delta v_h}=\rho v_h\,,\quad \frac{\delta \Hc}{\delta \chi}=-\nb\bcdot(\rho k^2\nb\chi)+[\rho,\Upsilon]\,,\label{func_der_5} \\
&&\frac{\delta \Hc}{\delta \Upsilon}=-\nb\bcdot(\rho \nb\Upsilon)+[\chi,\rho]\,,\quad \frac{\delta\Hc}{\delta\Omega}=\Lc^{-1}\frac{\delta\Hc}{\delta\chi}\,, \quad \frac{\delta\Hc}{\delta w}=\Delta^{-1}\frac{\delta\Hc}{\delta\Upsilon}\,. \label{func_der_6}
\end{eqnarray}
\subsection{Helically symmetric dynamics}
The helically symmetric dynamics is  described by means of the Hamiltonian \eqref{hs_hamiltonian} and the Poisson bracket \eqref{hs_poisson} as  
$\partial_t \eta_{_{HS}}=\{\eta_{_{HS}},\Hc\}^{^{XMHD}}_{_{HS}}$. Due to the helical symmetry and the compressibility,  the equations of motion appear much more involved than the corresponding equations of motion in \citet{Grasso2017}. For this reason we present here the dynamical equations for incompressible plasmas ($\rho=1$). Incompressible equations are obtained from the Hamiltonian and the Poisson bracket that correspond to $\rho=1$ and $w=0$, or equivalently by the compressible equations of motion by neglecting the dynamical equations for $\rho$ and $w$ and substituting $\Hc_w=0$ and $\Hc_\Omega=\chi$ in the rest (see Appendix \ref{app_A}),
%
\begin{eqnarray}
\partial_t v_h&&=k\left([\chi,k^{-1}v_h]+[k^{-1}B_h,\psi^*]\right)\,, \label{dyn_vh}\\
\partial_t\Omega&&=[\chi,\Omega]-2n\ell [\chi,k^3v_h] +[kv_h,k^{-1}v_h]\nn \\
&&+[k^{-1}B_h,k B_h^*]+[\Lc\psi,\psi^*]-2n\ell[k^3B_h,\psi^*]\,,\label{dyn_Omega} \\
\partial_t B_h^*&&= k^{-1}\big([\chi,kB_h^*]+[kv_h,\psi^*]-2n\ell k^4[\chi,\psi^*]
+d_i[kB_h^*,k^{-1}B_h]-d_i[\Lc\psi,\psi^*]\nn \\
&& -2n\ell d_ik^4[\psi^*,k^{-1}B_h]+2n\ell d_i[k^3B_h,\psi^*] +d_e^2[k^{-1}B_h,\Omega] -2n\ell d_e^2[k^{-1}B_h,k^3v_h]\nn\\
&&+d_e^2[\Lc\psi,k^{-1}v_h] -2n\ell d_e^2k^4[k^{-1}B_h,k^{-1}v_h]-2n\ell d_e^2[k^3B_h,k^{-1}v_h]\big)\,,\label{dyn_Bh}\\
\partial_t\psi^*&&=[\chi,\psi^*]+d_i[\psi^*,k^{-1}B_h]+d_e^2[k^{-1}B_h,k^{-1}v_h]\,. \label{dyn_psi}
\end{eqnarray}
Equations ~\eqref{dyn_vh}--\eqref{dyn_psi} differ from the corresponding dynamical equations of reference \citep{Grasso2017} owing to the presence of the scale factor $k$ and the purely helical terms with the coefficients $n\ell$. Setting $n=0$ we recover the equations of motion for incompressible, translationally symmetric plasmas, whereas for $\ell=0$ we restrict the motion to respect axial symmetry. 
\subsection{Bracket transformation}
In \citet{Lingam2015} the authors proved that the XMHD bracket \eqref{poisson} can be simplified to a form identical to the HMHD bracket by introducing a generalized vorticity variable
\begin{equation}
\bB^\pm=\bB^*+\gamma_{\pm}^{}\bnabla\times \bv \,.
\end{equation} 
This transformation was utilized in \citet{Grasso2017,Kaltsas2017} in order to simplify the bracket and the derivation of the symmetric families of Casimir invariants. For this reason we perform the transformation also for the helically symmetric bracket \eqref{hs_poisson}, rendering the subsequent analysis more tractable.  One can see that the corresponding scalar field variables, necessary for the poloidal representation of $\bB^\pm$, are connected to the variables $\eta_{_{HS}}$ as follows:
\begin{eqnarray}
B_h^\pm=B_h^*+\gamma_\pm (k^{-1}\Omega-2n\ell k^2v_h)\,, \label{Bh^pm} \\
\psi^{\pm}=\psi^*+\gamma_\pm k^{-1}v_h\,. \label{psi^pm}
\end{eqnarray}
Transformation of  the bracket requires expressing the functional derivatives in the new representation $(v_h,\chi,\Upsilon,B_h^{\pm},\psi^{\pm})$. Following an analogous procedure to that  employed in \citet{Lingam2015,Grasso2017,Kaltsas2017} we find 
\begin{eqnarray}
&& \bar{F}_{v_h} = F_{v_h}+\gamma_{\pm}k^{-1}F_{\psi^\pm}-2n\ell \gamma_\pm k^2 F_{B_h^\pm}\,,\label{hs_trans_1}\\
&&\bar{F}_\Omega= F_\Omega+\gamma_\pm k^{-1}F_{B_h^\pm}\,, \quad \bar{F}_w=F_w\,,   \label{hs_trans_2}\\
&& \bar{F}_{\psi^*}=F_{\psi^\pm}\,, \quad \bar{F}_{B_h^*}=F_{B_h^\pm}\,, \label{hs_trans_3}
\end{eqnarray}
where $\bar{F}$ denotes the functionals expressed in the original variable representation.  Upon inserting the transformation of  the functional derivatives  of  \eqref{hs_trans_1}--\eqref{hs_trans_3} into \eqref{hs_poisson} and expressing $B_h^*$ and $\psi^*$ in terms of $B_h^\pm$ and $\psi^\pm$ we obtain the following bracket:
\begin{eqnarray}
\{F,G\}^{^{XMHD}}_{_{HS}}&&=\int_Dd^3x\, \Big\{ F_\rho\Delta G_w-G_\rho\Delta F_w 
 +\rho^{-1}(\Omega-2n\ell k^3 v_h)\times \nn \\
 &&\times\big(
\left[F_{\Omega},G_{\Omega}\right] +k^{-2}[F_w,G_w] 
+\nb F_w \bcdot\nb G_\Omega- \nb F_\Omega\bcdot \nb G_w
\big)\nn\\
&&
+k^{-1}v_h\big(
[\rho^{-1}k F_{v_h},G_{\Omega}]-[\rho^{-1}k G_{v_h},F_\Omega] \nn \\
&&+ \nb \bcdot(\rho^{-1}kG_{v_h} \nb F_w)-\nb \bcdot(\rho^{-1}kF_{v_h} \nb G_w)
\big) \nn \\
&& +\rho^{-1}kB_h^\pm\Big([F_\Omega ,k^{-1}G_{B_h^\pm}] - [G_\Omega , k^{-1}F_{B_h^\pm}]\nn \\
&&+\nb F_w\bcdot\nb (k^{-1}G_{B_h^\pm})-\nb G_w\bcdot\nb(k^{-1} F_{B_h^\pm})
 \Big) \nn \\
&&+\psi^\pm\big(
[F_{\Omega},\rho^{-1}G_{\psi^\pm}]-[G_{\Omega},\rho^{-1} F_{\psi^\pm}] \nn \\
&&+[\rho^{-1}k F_{v_h},k^{-1}G_{B_h^\pm}]-[\rho^{-1}k G_{v_h},k^{-1}F_{B_h^\pm}] \nn \\
&& + \nb \bcdot(\rho^{-1}G_{\psi^\pm}\nb F_w) -  \nb \bcdot(\rho^{-1}F_{\psi^\pm}\nb G_w)
\big) \nn \\
&& -2n \ell \psi^{\pm}\big([F_\Omega,\rho^{-1}k^3 G_{B_h^\pm}]-[G_\Omega,\rho^{-1}k^3 F_{B_h^\pm}]\nn\\
&&+\nb\bcdot(\rho^{-1}k^3 G_{B_h^\pm}\nb F_w)-\nb\bcdot(\rho^{-1}k^3 F_{B_h^\pm}\nb G_w)\big)\nn \\
&&-\nu_{\pm} \rho^{-1}k B_h^\pm[k^{-1}F_{B_h^\pm},k^{-1}G_{B_h^\pm}]\nn \\
&&
- \nu_{\pm} \psi^\pm \left([\rho^{-1}F_{\psi^\pm},k^{-1}G_{B_h^\pm}]-[\rho^{-1}G_{\psi^\pm},k^{-1}F_{B_h^\pm}]\right) \nn\\
&& +2n\ell \nu_\pm \psi^{\pm}\left([\rho^{-1}k^3F_{B_h^\pm},k^{-1}G_{B_h^\pm}]-[\rho^{-1}k^3G_{B_h^\pm},k^{-1}F_{B_h^\pm}]\right)
\Big\} \, ,\label{hs_poisson_trans}
\end{eqnarray}
where $\nu_{\pm}:=d_i-2\gamma_\pm$. Note that the helically symmetric XMHD dynamics is described correctly by either using the parameter $\nu_+$ or the parameter $\nu_-$.
\section{Casimir invariants and equilibrium variational principle with helical symmetry}
\label{sec_IV}
As mentioned in Sec. \ref{sec_II}, the Casimir invariants are functionals that satisfy $\{F,\Cc\}=0$, $\forall F$.  For the bracket \eqref{hs_poisson_trans} this condition is equivalent to
\begin{eqnarray}
\int_Dd^3x\,\big(F_\rho \mathfrak{R}_1+F_{w}\mathfrak{R}_2+\rho^{-1}kF_{v_h}\mathfrak{R}_3+F_\Omega\mathfrak{R}_4 +k^{-1}F_{B_h^\pm}\mathfrak{R}_5+\rho^{-1}F_{\psi^\pm}\mathfrak{R}_6
\big)=0\, ,
 \label{cas_det_0}
\end{eqnarray}
where $\mathfrak{R}_i\,, i=1,...,6$ are expressions obtained by manipulating  the bracket $\{F,\Cc\}$ so as to extract as common factors the functional derivatives of the arbitrary functional $F$. Requiring \eqref{cas_det_0} to be satisfied for arbitrary variations is equivalent  to the independent vanishing of the expressions $\mathfrak{R}_i$, i.e., 
\begin{equation}
\mathfrak{R}_i=0\,, \qquad  i=1,2,...,6\, . \label{cas_det_sys}
\end{equation}
The expressions for the $\mathfrak{R}_i\,, i=1,...6$, read 
\begin{eqnarray}
\mathfrak{R}_1&&=\Delta \Cc_w=\Cc_\Upsilon \,,\label{cas_det_1}\\
\mathfrak{R}_2&&=-\Delta \Cc_\rho-[\rho^{-1}k^{-2}\Omega,\Cc_w]+2n\ell [\rho^{-1}kv_h,\Cc_w] \nn \\
&&+\nb\bcdot (\rho^{-1}\Cc_{\psi^\pm}\nb\psi^\pm)-2n\ell\nb\bcdot (\rho^{-1}k^{3}\Cc_{B_h^\pm}\nb\psi^\pm)+\nb\bcdot(\rho^{-1}k\Cc_{v_h}\nb(k^{-1}v_h)) \nn \\
&&-\nb\bcdot(\rho^{-1}\Omega\nb\Cc_\Omega)+2n\ell \nb\bcdot(\rho^{-1}k^3v_h\nb \Cc_\Omega)-\nb\bcdot(\rho^{-1}k B_h^\pm\nb (k^{-1}\Cc_{B_h^\pm}))\,,
 \label{cas_det_2} \\ 
\mathfrak{R}_3&&=[\Cc_\Omega,k^{-1}v_h]+\bnabla (k^{-1}v_h)\bcdot\nb \Cc_w-[\psi_\pm,k^{-1}\Cc_{B_h^\pm}]\,, \label{cas_det_3}\\
\mathfrak{R}_4&&=\nb\bcdot (\rho^{-1}\Omega\nb\Cc_w)-2n\ell \nb\bcdot(\rho^{-1}k^3v_h\nb\Cc_w)\nn\\
&&- [\rho^{-1}\Omega,\Cc_{\Omega}]+2n\ell [\rho^{-1}k^3v_h,\Cc_\Omega] -[k^{-1}v_h,\rho^{-1}k\Cc_{v_h}] \nn \\
&&-[\psi^\pm,\rho^{-1}\Cc_{\psi^\pm}]-[\rho^{-1}kB_h^\pm,k^{-1}\Cc_{B_h^\pm}]+2n\ell [\psi^\pm,\rho^{-1}k^3\Cc_{B_h^\pm}] \,,\label{cas_det_4}
\\
\mathfrak{R}_5&&=[\rho^{-1}k\Cc_{v_h},\psi^\pm]+[\Cc_\Omega,\rho^{-1}kB_h^\pm] +\nb\bcdot (\rho^{-1}kB_h^\pm\nb \Cc_w)-2n\ell \rho^{-1}k^4[\Cc_\Omega,\psi^\pm]
\nn\\
&&-2n\ell \rho^{-1}k^4\nb\psi^\pm\bcdot\nb\Cc_w+\nu_{\pm} [\psi_\pm,\rho^{-1}\Cc_{\psi_\pm}]+\nu_{\pm} [\rho^{-1}kB_h^\pm,k^{-1}\Cc_{B_h^\pm}]\nn\\
&&-2n\ell\nu_\pm\rho^{-1}k^4[\psi^\pm,k^{-1}\Cc_{B_h^\pm}]+2n\ell \nu_\pm [\rho^{-1}k^3\Cc_{B_h^\pm},\psi^\pm]\,,\label{cas_det_5}\\
\mathfrak{R}_6&&= [\Cc_\Omega,\psi_\pm]+\nb \psi_\pm\bcdot \nb \Cc_w+\nu_{\pm} [\psi_\pm,k^{-1}\Cc_{B_h^\pm}] \, .\label{cas_det_6}
\end{eqnarray}
Equation $\mathfrak{R}_1=0$, i.e. $\Cc_\Upsilon=0$, implies that the Casimirs are independent of $\Upsilon$. 
We observe that  \eqref{cas_det_sys} are satisfied automatically for $\Cc_\rho$=const.,  which amounts to the conservation of mass density, 
\begin{equation}
\Cc_m=\int_Dd^3x\, \rho \,.
 \label{mass_cas}
\end{equation}
For the rest of the Casimirs we follow a similar procedure as in Sec.~IIIB of our previous study \citep{Kaltsas2017}. Although the analysis is now more involved due to the purely helical terms appearing in \eqref{cas_det_2}--\eqref{cas_det_6}, it turns out that it is  not difficult to make the necessary adaptions for computing the helically symmetric Casimirs. For this reason we avoid presenting the  procedure once again, instead giving directly the resulting Casimir invariants, which in terms of the original magnetic field variables $(B_h^*,\psi^*)$ are given by
\begin{eqnarray}
\Cc_1&=&\int_Dd^3x\,\Big[(kB^*_h+\gamma\Omega-2n\ell \gamma k^3 v_h)\Fc(\psi^*+\gamma k^{-1} v_h)\nn\\&&\hspace{1cm}+2n\ell k^4 \tilde{\Fc}(\psi^*+\gamma k^{-1} v_h)\Big]\,,\label{hs_cas_1}\\
\Cc_2&=&\int_Dd^3x\,\Big[(kB^*_h+\mu\Omega-2n\ell \mu k^3 v_h)\Gc(\psi^*+\mu k^{-1} v_h)\nn \\
&&\hspace{1cm}+2n\ell k^4 \tilde{\Gc}(\psi^*+\mu k^{-1} v_h)\Big]\,, \label{hs_cas_2}\\
\Cc_3&=&\int_Dd^3x\, \rho \Mc (\psi^*+\gamma k^{-1} v_h)\,, \label{hs_cas_3} \\
\Cc_4&=&\int_Dd^3x\, \rho \Nc(\psi^*+\mu k^{-1}v_h) \label{hs_cas_4}\,,
\end{eqnarray} 
where the parameters $\gamma$ and $\mu$ are $(\gamma,\mu)=(\gamma_+,\gamma_-)$,  $\Fc$, $\Gc$, $\Mc$, $\Nc$ are arbitrary functions and $\tilde{\Fc}$, $\tilde{\Gc}$ are defined by
\begin{equation}
\tilde{\Fc}=\int_0^{\psi^*+\gamma k^{-1}v_h}\Fc(g)dg\,,  \qquad \tilde{\Gc}=\int_0^{\psi^*+\mu k^{-1}v_h}\Gc(g)dg\,.
\end{equation}
Obviously $\Cc_m$ is just a special case of the functionals $\Cc_3$, $\Cc_4$.  Upon substituting the functionals \eqref{hs_cas_1}--\eqref{hs_cas_4} into \eqref{cas_det_1}--\eqref{cas_det_6},  we can verify that \eqref{cas_det_sys} are satisfied and thus $\Cc_1,\,\Cc_2,\, \Cc_3$ and $\Cc_4$ are indeed conserved quantities of the helically symmetric XMHD. The interesting new feature of these Casimirs is the presence of two purely helical terms appearing in $\Cc_1$ and $\Cc_2$, which vanish for either $n=0$ or $\ell=0$. An analogous helical term, that depend on $\psi$, having coefficient $2n\ell$, appears also in the Casimirs of ordinary MHD \citep{Andreussi2012}. In the case of XMHD the helical terms depend on $\psi^*$ and on the  helical velocity $v_h$,  this additional dependence on $v_h$ emerges due to the presence of the vorticity in \eqref{cas_xmhd}.
\subsection{MHD limit}
\label{sec_MHDlim}
In \citet{Kaltsas2017} various limits of the symmetric XMHD Casimirs to  those of the simpler models of Hall MHD, ordinary MHD,  and inertial MHD were obtained. Here,  to corroborate  that the computed invariants are correct,  we take the MHD limit,  anticipating the recovery of  the invariants found in \citet{Andreussi2012}.
For the MHD limit we  set $d_e=0$ (Hall MHD) and $d_i=0$. If we set only $d_e=0$ we exclude electron inertial contributions and we obtain the Hall MHD Casimirs
\begin{eqnarray}
\Cc_1^{^{HMHD}}&=&\int_Dd^3x\,\Big[(kB_h+d_i\Omega-2n\ell d_i k^3 v_h)\Fc(\psi+d_i k^{-1} v_h)\nn \\
&&\hspace{1cm}+2n\ell k^4 \tilde{\Fc}(\psi+d_i k^{-1} v_h)\Big]\,,\label{hall_cas_1}\\
\Cc_2^{^{HMHD}}&=&\int_Dd^3x\,\left[kB_h\Gc(\psi)+2n\ell k^4 \tilde{\Gc}(\psi)\right]\,, \label{hall_cas_2}\\
\Cc_3^{^{HMHD}}&=&\int_Dd^3x\, \rho \Mc (\psi+d_i k^{-1} v_h)\,, \label{hall_cas_3} \\
\Cc_4^{^{HMHD}}&=&\int_Dd^3x\, \rho \Nc(\psi) \label{hall_cas_4}\,.
\end{eqnarray}
For the corresponding MHD families of invariants we additionally require $d_i\rightarrow 0$ in \eqref{hall_cas_1}--\eqref{hall_cas_4}. From the resulting set of Casimirs the cross-helicity and the helical momentum are absent. This is a characteristic peculiarity, encountered when the MHD limit of models with Hall physics contributions is considered (e.g.\  see \citet{Kaltsas2017,Hazeltine1987,Abdelhamid2015,Yoshida2013}). In \citet{Kaltsas2017} we resolved this peculiarity by expanding the invariants $\Cc_1^{^{HMHD}}$, $\Cc_3^{^{HMHD}}$ about $\psi$, then by rescaling the arbitrary functions we managed to show that, since the terms that diverge when $d_i\rightarrow 0$ are already Casimirs, the rest of the terms translate to the MHD Casimirs. Doing so for the helically symmetric Casimirs we arrive at
\begin{eqnarray}
\Cc_1^{^{MHD}}&=&\int_Dd^3x\,\left[B_hv_h\Fc'(\psi)+\Omega \Fc(\psi)\right]\,,\label{mhd_cas_1}\\
\Cc_2^{^{MHD}}&=&\int_Dd^3x\,\left[kB_h\Gc(\psi)+2n\ell k^4 \tilde{\Gc}(\psi)\right]\,, \label{mhd_cas_2}\\
\Cc_3^{^{MHD}}&=&\int_Dd^3x\, \rho k^{-1}v_h\Mc (\psi)\,, \label{mhd_cas_3} \\
\Cc_4^{^{MHD}}&=&\int_Dd^3x\, \rho \Nc(\psi) \label{mhd_cas_4}\,.
\end{eqnarray}
The functionals \eqref{mhd_cas_1}--\eqref{mhd_cas_4} are indeed the correct helically symmetric MHD Casimir invariants \citep{Andreussi2012}.
\subsection{Equilibrium variational principle with helical symmetry}
With the helically symmetric Casimirs at hand, we can build the EC variational principle to obtain equilibrium conditions. For analogous utilizations of this methodology for symmetric or 2D plasmas the reader is referred to \citet{Holm1985,Almaguer1988,Andreussi2008,Tassi2008,Andreussi2010,Andreussi2012,Moawad2013,
Morrison2014,Kaltsas2017}. As mentioned in Sec.~\ref{sec_II}, the EC principle states that the phase space points that nullify the first variation EC functional $\mathfrak{F}$  are equilibrium points. In our case requiring the vanishing of $\delta\mathfrak{F}$ amounts to
 \begin{eqnarray}
&&\delta\int_Dd^3x\, \bigg\{\rho\left(\frac{v_h^2}{2}+\frac{k^2}{2}|\bnabla \chi|^2+
[\Upsilon,\chi]+\frac{|\nb\Upsilon|^2}{2}+U(\rho)\right) \nn \\
&& \hspace{0.5cm}+\frac{B_h^*B_h}{2}+\frac{k^2}{2}\nb\psi^*\bcdot\nb\psi 
 -(kB^*_h+\gamma\Omega-2n\ell \gamma k^3 v_h)\Fc(\varphi)-2n\ell k^4 \tilde{\Fc}(\varphi) 
 \nn\\
&&\hspace{0.5cm}
-(kB^*_h+\mu\Omega-2n\ell \mu k^3 v_h)\Gc(\xi)-2n\ell k^4 \tilde{\Gc}(\xi)- \rho \Mc(\varphi)-\rho \Nc(\xi)\bigg\}=0\,,
 \label{var_pri_main}
\end{eqnarray}
where $\varphi:=\psi^*+\gamma k^{-1}v_h$, $\xi:=\psi^*+\mu k^{-1}v_h$ and 
\begin{equation}
\tilde{\Fc}:=\int^{\varphi} \Fc(g)dg \quad \mathrm{and}\quad  \tilde{\Gc}:=\int^{\xi}  \Gc(g)dg\,.
\end{equation}
Since the variations of the field variables are independent, \eqref{var_pri_main} is satisfied if the coefficients of the variations of the field variables  vanish. This requirement, upon exploiting the relations \eqref{func_der_1}--\eqref{func_der_6}, leads to the following equilibrium conditions:
\begin{eqnarray}
\delta\rho &:&  \left[\rho U(\rho)\right]_\rho+\frac{v^2}{2}-\Mc(\varphi)-\Nc(\xi)
+\frac{d_e^2}{2\rho^2}\left(J_h^2+k^2|\nb (k^{-1}B_h)|^2\right)=0\,,
 \label{d_rho}\\
 \delta \Upsilon &:&  -\nb\bcdot(\rho\nb\Upsilon)+[\chi,\rho]=0\,, \label{d_Y}\\
 \delta\chi &:&  -\nb\bcdot( \rho k^2 \nb \chi)+[\rho,\Upsilon]-\gamma \Lc \Fc(\varphi)-\mu \Lc \Gc(\xi)=0 \,,\label{d_chi}\\
\delta v_h &:&   \rho v_h-\rho k^{-1}\left[\gamma\Mc '(\varphi)+\mu \Nc '(\xi)\right] -B_h^*\left[\gamma\Fc'(\varphi)+\mu\Gc'(\xi)\right]
\nn\\
&&\hspace{0.4cm}-k^{-1}(\Omega-2n\ell k^3v_h)\left[\gamma^2 \Fc'(\varphi)+\mu^2 \Gc'(\xi)\right]=0 \,,
\label{d_vh} \\
\delta B_h^*&:&  B_h-k\left[\Fc(\varphi)+ \Gc(\xi)\right]=0 \,,\label{d_Bh*}\\
\delta \psi^* &:& \Lc \psi - kB_h^*\left[\Fc'(\varphi)+\Gc'(\xi)\right]-2n\ell k^4 \left[\Fc(\varphi)+\Gc(\xi)\right]\nn\\
&&\hspace{0.4cm}-(\Omega-2n\ell k^3v_h)\left[\gamma\Fc'(\varphi)+\mu\Gc'(\xi)\right]-\rho\left[\Mc'(\varphi)+\Nc'(\xi)\right]=0\,.
 \label{d_psi*}
\end{eqnarray}
Note that the lhs of \eqref{d_rho}--\eqref{d_psi*} are the coefficients of the variations $(\delta\rho,\delta \Upsilon,\delta\chi, \delta v_h,\delta B_h^*,\delta\psi^*)$ in $\delta\mathfrak{F}$. In addition to these terms, some surface boundary terms emerged in $\delta \mathfrak{F}$, due to integration by parts. We assumed that those terms vanish, which is true if the variations $\delta\Upsilon, \delta\chi, \delta\psi^*$ vanish on the boundary $\partial D$. The first equation \eqref{d_rho} represents a Bernoulli law
\begin{eqnarray}
\tilde{p}(\rho)&=&\rho \left[\Mc(\varphi)+\Nc(\xi)\right]-\rho \frac{v^2}{2}
-\frac{d_e^2}{2\rho}\left[J_h^2+k^2|\nb (k^{-1}B_h)|^2\right]\,,
 \label{bernoulli_gen}
\end{eqnarray}
where $\tilde{p}:=\rho [\rho U(\rho)]_\rho=\rho h(\rho)$ where $h(\rho)$ is the total enthalpy ($\tilde{p}={\Gamma} p/({\Gamma-1})$ if we adopt the equation of state $p \propto \rho^\Gamma$ with $\Gamma$ being the adiabatic constant). It describes the effect of  macroscopic equilibrium flow including the electron inertial effect, expressed via the magnetic terms, in the total plasma pressure. The rest of the equations can be cast into a Grad-Shafranov or a JFKO system as in the case with translational symmetry.
\subsection{The JFKO-Bernoulli system}
The system \eqref{d_rho}--\eqref{d_psi*} can be cast in a JFKO-Bernoulli PDE form that describes completely helically symmetric XMHD equilibria. This can be done by exploiting \eqref{d_Y}, \eqref{d_chi}, \eqref{d_Bh*} and \eqref{B_h*} in order to turn \eqref{d_vh} and \eqref{d_psi*} into a coupled system for the flux functions $\varphi$ and $\xi$. These equations, except of their coupling to the Bernoulli equation, are additionally coupled to the definition \eqref{psi*} given in terms of $\varphi$ and $\xi$ 
expressing essentially the Ampere's law. The derivation of the system requires tedious algebraic manipulations that we omit  here; however,  the steps are analogous to those used in \citet{Kaltsas2017} for the derivation of the corresponding system. The JFKO equations for barotropic XMHD are   
\begin{eqnarray}
&&\hspace{-0.5cm}(\gamma^2+d_e^2)\Fc'\nb\bcdot\left(\frac{k^2}{\rho}\nb\Fc\right)=(1+s)k^2(\Fc+\Gc)\Fc'+\rho \Mc'+\bigg(\frac{\mu}{\gamma-\mu}-2n\ell \frac{d_e^2}{\rho}k^2\Fc' \bigg)\Lc\psi\nn\\
&&\hspace{2cm}-2n\ell\frac{\mu}{\gamma-\mu}k^4(\Fc+\Gc)- k^2\left[\frac{\rho}{(\gamma-\mu)^2}+2n\ell\frac{\gamma}{\gamma-\mu}k^2\Fc'\right](\varphi-\xi)\,,\label{jfko_1} \\
&&\hspace{-0.5cm}(\mu^2+d_e^2)\Gc'\nb\bcdot\left(\frac{k^2}{\rho}\nb\Gc\right)=(1+s)k^2(\Fc+\Gc)\Gc'+\rho \Nc'-\bigg(\frac{\gamma}{\gamma-\mu}+2n\ell \frac{d_e^2}{\rho}k^2\Gc' \bigg)\Lc\psi\nn\\
&&\hspace{2cm}+2n\ell\frac{\gamma}{\gamma-\mu}k^4(\Fc+\Gc)+k^2\left[\frac{\rho}{(\gamma-\mu)^2}-2n\ell\frac{\mu}{\gamma-\mu}k^2\Gc'\right](\varphi-\xi)\,,\label{jfko_2}\\
&&\Lc\psi=k^2\frac{\rho}{d_e^2}\left[\frac{\mu\varphi-\gamma \xi}{\mu-\gamma}-\psi+2n\ell d_e^2\rho^{-1}k^2(\Fc+\Gc)\right]\,, \label{jfko_3}
\end{eqnarray}
where $s:=4n^2\ell^2d_e^2\rho^{-1}k^4$.
The equations above coupled to the Bernoulli equation \eqref{bernoulli_gen} describe completely the equilibria in terms of the flux functions $\psi$, $\varphi$, $\xi$ and of the density $\rho$, for given free functions $\Fc(\varphi)$, $\Gc(\xi)$, $\Mc(\varphi)$, $\Nc(\xi)$ and a thermodynamic closure $p=p(\rho)$, since all physical quantities of interest can be expressed in terms of $\psi$, $\varphi$, $\xi$ and $\rho$. Namely, the helical component of the flow is 
\begin{equation}
v_h=k\frac{\varphi-\xi}{\gamma-\mu}\,, \label{v_h}
\end{equation}
the helical magnetic field is given by \eqref{d_Bh*}, the poloidal field is simply $\nb\psi\times\bh$,  while for the poloidal velocity we need to observe that \eqref{d_Y} and \eqref{d_chi} can be written as 
\begin{equation}
\bh\bcdot\nb\times \boldsymbol{Q}=0  \quad \mathrm{and}\quad \nb\bcdot (k^2\boldsymbol{Q})=0 \,,
\end{equation}
with 
\begin{equation}
\boldsymbol{Q}:=\rho \nb\chi-\rho k^{-2}\nb\Upsilon\times\bh-\gamma \nb \Fc-\mu \nb \Gc\,.\label{Q}
\end{equation}
Therefore the mutual solution of \eqref{d_Y} and \eqref{d_chi}, should satisfy
\begin{equation}
\rho \nb\chi-\rho k^{-2} \nb \Upsilon\times \bh=\gamma  \nb \Fc+\mu \nb \Gc \,. \label{Q_eq}
\end{equation}
Upon taking the cross product of \eqref{Q_eq} with $\bh$, we obtain the poloidal velocity 
\begin{equation}
\bv_p=\rho^{-1}\left(\gamma\nb\Fc+\mu\bnabla\Gc\right)\times\bh\,.\label{v_p}
\end{equation} 
Note that \eqref{Q_eq} was necessary in the derivation of the equations \eqref{jfko_1}--\eqref{jfko_2}. 
Due to the number of the PDEs that have to be solved simultaneously and  the consequence of the symmetry that inserts additional terms and the strong coupling among the equations, the solution of this system is not trivial even in the context of numerical computing. For this reason we present below special cases of equilibria, including axisymmetric XMHD, incompressible XMHD, barotropic and incompressible Hall MHD equilibria with helical symmetry. To our knowledge all these reduced kinds of equilibria have not been studied so far. In the next section we present the corresponding system of Grad-Shafranov or JFKO equations for each of the aforementioned equilibria. 
\section{Special Equilibria}
\label{sec_V}
\subsection{Axisymmetric barotropic XMHD}
The axisymmetric equilibrium  equations are obtained by setting  the helical angle $a$ to   zero, i.e.,  $\ell=0$ and $n=-1$, so the parameter $s$ is zero and the scale factor $k=1/r$ and $\bh=r^{-1}\hat{e}_\phi$. With these parameters,  \eqref{jfko_1}--\eqref{jfko_3} reduce to the following Grad-Shafranov system:
\begin{eqnarray}
&&\hspace{-0.5cm}(\gamma^2+d_e^2)\Fc'r^2\nb\bcdot\left(\frac{\Fc'}{\rho}\frac{\nb\varphi}{r^2}\right)=\Fc'(\Fc+\Gc)+r^2\rho \Mc'-\frac{\mu}{\gamma-\mu}\Delta^*\psi -\rho \frac{\varphi-\xi}{(\gamma-\mu)^2}\,, \label{as_gs_1} \\
&&\hspace{-0.5cm}(\mu^2+d_e^2)\Gc' r^2 \nb\bcdot\left(\frac{\Gc'}{\rho} \frac{\nb \xi}{r^2} \right)=\Gc' (\Fc+\Gc)+r^2 \rho \Nc' +\frac{\gamma}{\gamma-\mu}\Delta^*\psi +\rho \frac{\varphi-\xi}{(\gamma-\mu)^2}\,, \label{as_gs_2} \\
&&\Delta^*\psi=\frac{\rho}{d_e^2}\left(\psi-\frac{\mu \varphi-\gamma \xi}{\mu-\gamma}\right)\,, \label{as_gs_3} 
\end{eqnarray}
where $\Delta^*:=r^2\nb\bcdot\left(\nb/r^2\right)$ is the so-called Shafranov operator. The Bernoulli equation \eqref{bernoulli_gen} assumes the form:
\begin{eqnarray}
\tilde{p}(\rho)&=&\rho \left[\Mc(\varphi)+\Nc(\xi)\right]-\rho \frac{v^2}{2}
-\frac{d_e^2}{2\rho}\left[J_\phi^2+r^{-2}|\nb (rB_\phi)|^2\right]\,,
 \label{bernoulli_gen_ax}
\end{eqnarray}
where $J_\phi=-r^{-1}\Delta^*\psi $ is the toroidal current density.  For $d_e=0$ we obtain the axisymmetric Hall MHD Grad-Shafranov-Bernoulli system \citep{Throumoulopoulos2006}.
\subsection{Incompressible equilibria}
To obtain the equilibrium  system for incompressible plasmas with uniform mass density, we set $\rho=1$. Note that   incompressibility may refer also to the kind of the flows, i.e., flows with divergence-free velocity  fields that renders the mass density  a Lagrangian invariant, that is, $\rho$ is advected by the flow. Here we address the simpler case where the mass density is constant. One should be careful when adopting this assumption because it has to be imposed a priori, i.e.,  before varying the EC functional. This is because, if we use the barotropic version of the EC functional to derive the equilibrium equations and then impose the uniformness of the mass density, this  will result in a restricted class of equilibria because  the Bernoulli equation \eqref{bernoulli_gen} will act as an additional constraint on the permissible equilibria. However  for uniform mass density,  no Bernoulli equation occurs via the variational principle and the computation of the pressure decouples from the PDE problem. Ultimately the resulting equilibrium equations will be given by \eqref{d_chi}--\eqref{d_psi*} with $\rho=1$. This system   leads to the equilibrium system of  \eqref{jfko_1}--\eqref{jfko_3} with $\rho=1$, that is,  the differential operators on the lhs of \eqref{jfko_1} and \eqref{jfko_2}  reduce to the operator $-\Lc$ acting on $\Fc$ and $\Gc$,  respectively.
Those equations can alternatively be derived directly by taking projections of the starting stationary XMHD equations. We have verified that this method leads to the same JFKO system.
The pressure can be computed from \eqref{mom_eq} by setting $\partial_t\bv=0$, taking the divergence of the resulting equation and acting with the inverse of the Laplacian operator in order to solve for the pressure, leading to the following equation:
\begin{equation}
p=\Delta^{-1}\nb\bcdot\left(\bv\times \nb\times\bv+\bJ\times\bB^*\right)-\frac{v^2}{2}-\frac{d_e^2}{2}J^2\,.\label{p_inco_0}
\end{equation}
If we employ the helically symmetric representation \eqref{hel_sym_magn_field}, \eqref{hel_sym_vel_field} for the fields $\bB^*$, $\bv$ and $\bB$ and use the equilibrium equations \eqref{d_chi}-\eqref{d_psi*} with $\rho=1$, then we can prove that 
\begin{equation}
\bv\times \nb\times\bv+\bJ\times\bB^*=\nb\Mc(\varphi)+\nb\Nc(\xi)\,, \label{equi_eq_inco}
\end{equation}
so from \eqref{p_inco_0} and \eqref{equi_eq_inco}, we deduce that the incompressible pressure is given by 
\begin{equation}
p=\Mc(\varphi)+\Nc(\xi)-\frac{v^2}{2}-\frac{d_e^2}{2}J^2\,.\label{p_inco}
\end{equation}
\subsection{Hall MHD equilibria}
The Hall MHD limit is effected by setting $d_e=0$ and thereby neglecting  electron inertial effects.  Thus,  $\gamma=d_i$, $\mu=0$,  and the flux functions become  $\varphi=\psi+d_i k^{-1}v_h$ and  $\xi=\psi$. In this model, only ion drift effects are considered and the electron surfaces coincide with the magnetic field surfaces. The JFKO system for computing  the poloidal ion and magnetic fluxes is 
\begin{eqnarray}
&&d_i^2\Fc'\nb\bcdot\left(\frac{k^2}{\rho}\nb\Fc\right)=k^2(\Fc+\Gc)\Fc'+\rho \Mc'- k^2\left[\frac{\rho}{d_i^2}+2n\ell k^2\Fc'\right](\varphi-\psi)\,,\label{hall_jfko_1} \\
&&\Lc\psi=k^2(\Fc+\Gc)\Gc'+\rho \Nc'+2n\ell k^4(\Fc+\Gc)+k^2\rho\frac{(\varphi-\psi)}{d_i^2}\,.\label{hall_jfko_2}
\end{eqnarray}
These equilibria are completely determined through the coupling of the equations above with a Bernoulli law, which can be deduced from \eqref{bernoulli_gen} for $d_e=0$, allowing the computation of  the mass density $\rho$ self-consistently given an equation of state $P(\rho)$. 
So the HMHD Bernoulli equation is simply
\begin{equation}
\tilde{p}(\rho)=\rho\left[\Mc+\Nc-k^2\frac{(\varphi-\psi)^2}{2d_i^2}\right]-d_i^2 k^2\frac{(\Fc')^2}{2\rho}|\nb\varphi|^2\,. \label{bernoulli_hall}
\end{equation}
Also from \eqref{v_h} and \eqref{v_p} we have  
\begin{equation}
v_h=k\frac{\varphi-\psi}{d_i}
\quad \mathrm{and}\quad  
\bv_p=d_i \frac{\Fc'}{\rho} \nb\varphi\times\bh\,.
\end{equation} 
For $\ell=0$, \eqref{hall_jfko_1}, \eqref{hall_jfko_2} and \eqref{bernoulli_hall} reduce to the axisymmetric Grad-Shafranov-Bernoulli system of \citep{Throumoulopoulos2006}. For the baroclinic version of the axisymmetric HMHD equilibrium equations the reader is referred to \citet{Hameiri2013,Guazzotto2015}.
\subsection{Incompressible HMHD equilibria and the Double-Beltrami solutions}
\begin{figure}
\includegraphics[scale=0.35]{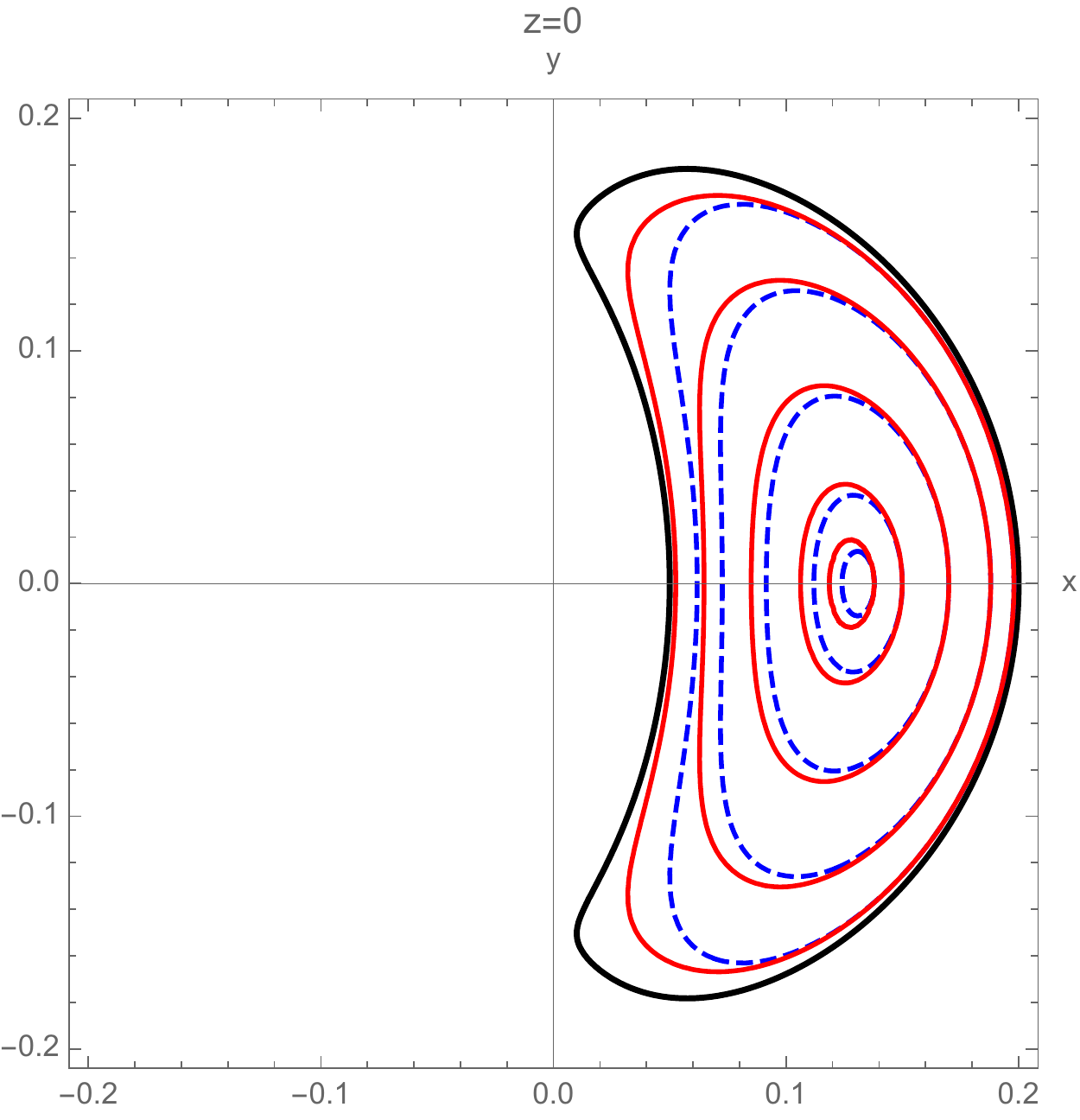}\includegraphics[scale=0.35]{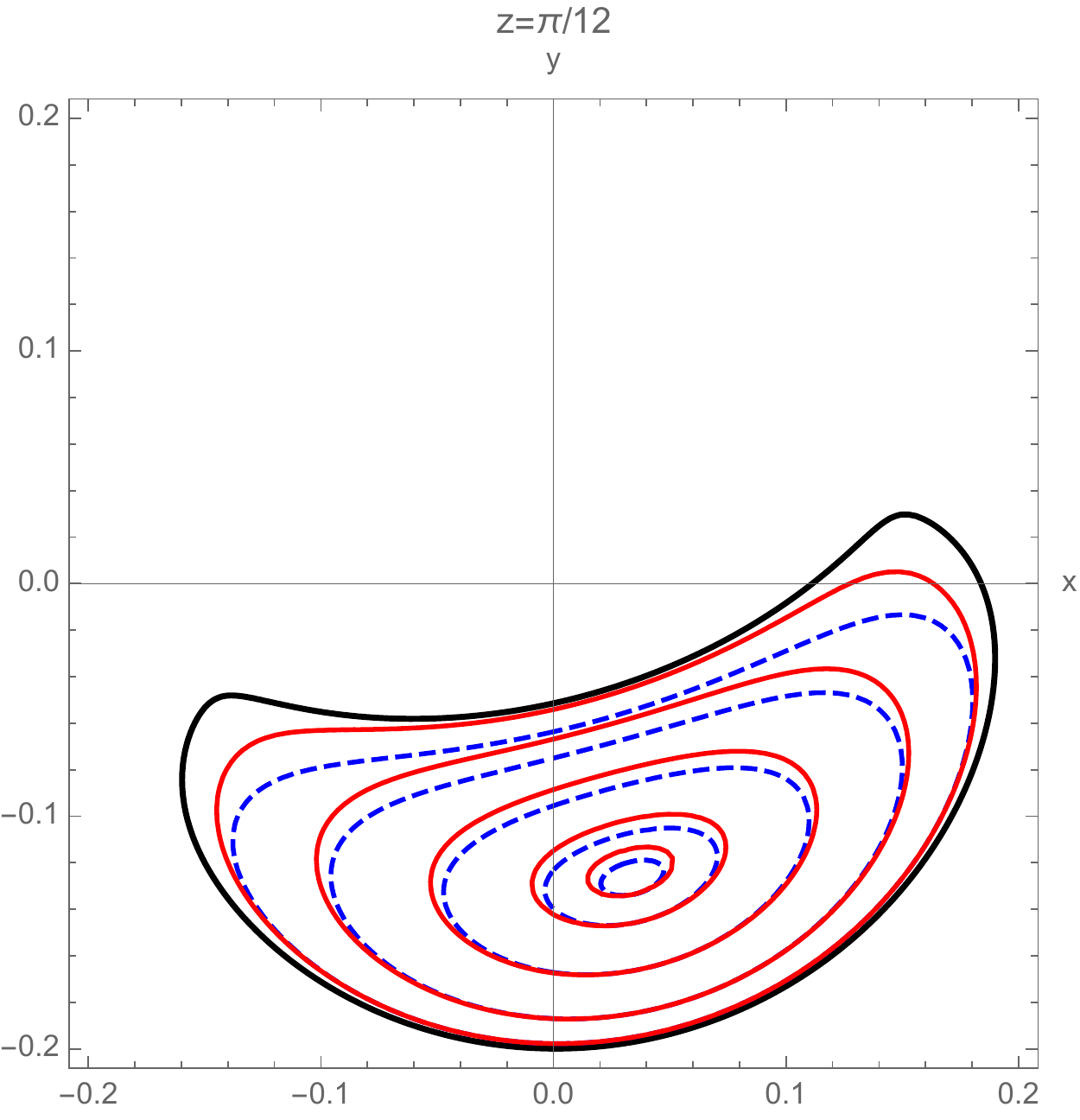}\includegraphics[scale=0.35]{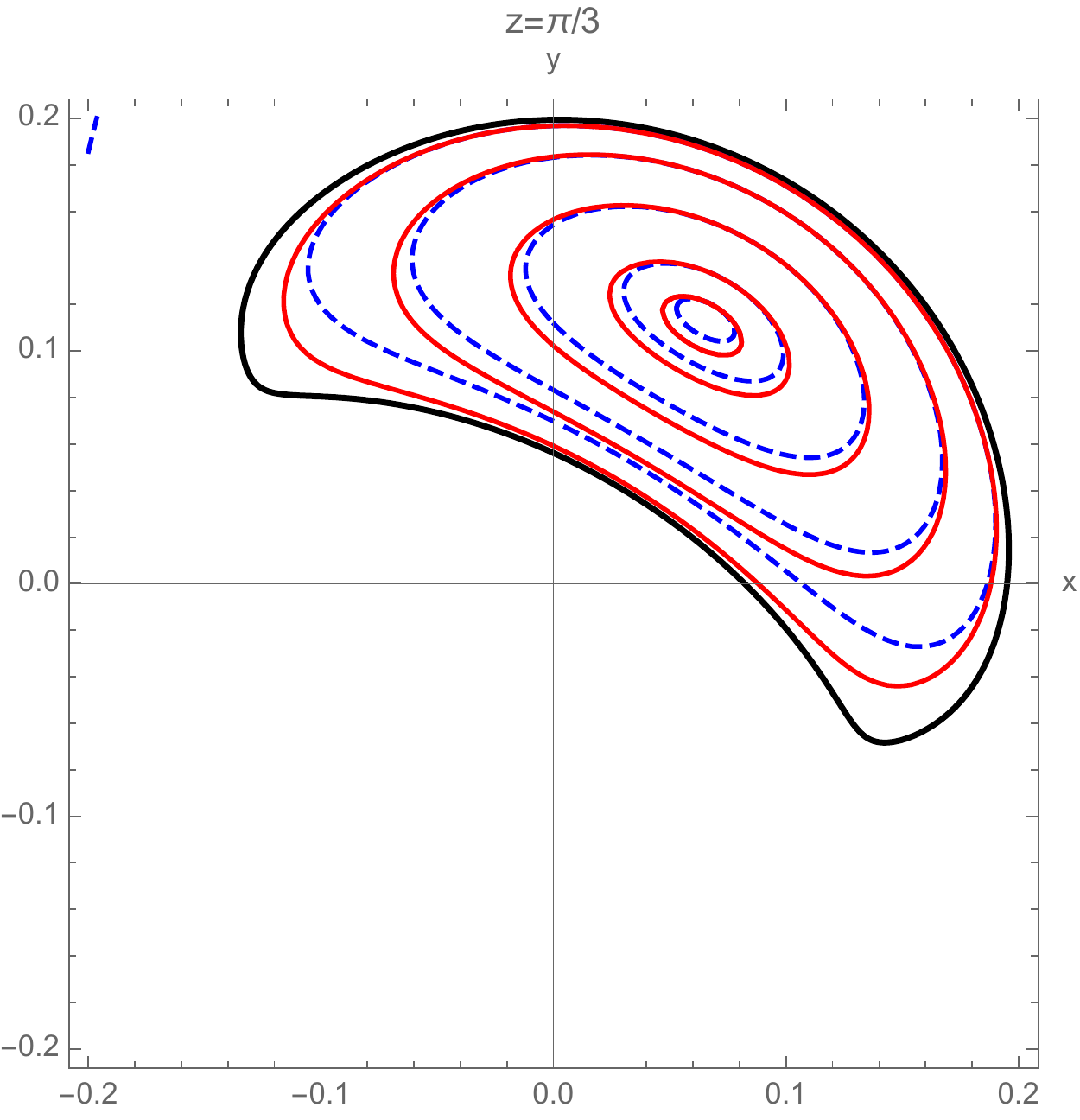}
\caption{The magnetic (solid red) and the ion (dashed blue) surfaces of the analytic DB equilibria with helical symmetry in connection with \eqref{db_flux} and \eqref{psi_beltrami} in three different sections, namely $z=0$, $z=\pi/12$, $z=\pi/3$. The parameters $\ell$ and $n$ are $\ell=1$ and $n=5$ corresponding to five helical windings for distance $2\pi$ covered in the $z$-direction. The contours have been plotted on the $(x,y)$ plane (perpendicular to the $z$-direction). }
\label{fig_1}
\end{figure}

As in the case of XMHD, to obtain an equilibrium system that is not constrained by an equation arising from  the $\rho$ functional derivative of the EC functional, we   take $\rho=1$ before varying the EC functional. If we do so for the HMHD model then the equilibrium system reduces to \eqref{hall_jfko_1}--\eqref{hall_jfko_2} with $\rho=1$, i.e.,  we have 
 \begin{eqnarray}
d_i^2\Fc'\Lc\Fc&=&-k^2(\Fc+\Gc)\Fc'- \Mc'+ k^2\left(d_i^{-2}+2n\ell k^2\Fc'\right)(\varphi-\psi)\,,\label{inco_hall_jfko_1} \\
\Lc\psi&=&k^2(\Fc+\Gc)\Gc'+\Nc'+2n\ell k^4(\Fc+\Gc)+k^2 {(\varphi-\psi)}{d_i^{-2}}\,.
\label{inco_hall_jfko_2}
\end{eqnarray}
The pressure can be computed using \eqref{p_inco} with $d_e=0$. To obtain solutions for the fluxes $\varphi$ and $\psi$, we need to specify the free functions $\Fc$, $\Gc$, $\Mc$ and $\Nc$. There exist a particular Ansatz for the free functions, for which the system \eqref{inco_hall_jfko_1}--\eqref{inco_hall_jfko_2} assumes an analytic solution. In this case the magnetic and velocity fields are superpositions of two Beltrami fields and the functions $\varphi$ and $\psi$ are expressed as linear combinations of the corresponding poloidal flux functions of the Beltrami fields. 
The generic linear Ansatz, for the system \eqref{inco_hall_jfko_1}--\eqref{inco_hall_jfko_2} is 
\begin{equation}
\Fc=f_0+f_1\varphi\,,\quad \Gc=g_0+g_1\psi\,,\quad \Mc=m_0+m_1\varphi\,,\quad \Nc=n_0+n_1\psi\,,\label{ansatz}
\end{equation}
where $f_0,\, f_1,\, g_0,\, g_1,\, m_0,\, n_0$ are constant parameters, leads to the following equations for helically symmetric HMHD equilibria:
\begin{equation}
k^{-2}\Lc \begin{pmatrix}
\varphi \\ \psi
\end{pmatrix} =\begin{pmatrix}
\Wc_1 & \Wc_2\\
\Wc_3 & \Wc_4
\end{pmatrix}\begin{pmatrix}
\varphi \\ \psi
\end{pmatrix}+\begin{pmatrix}
\Rc_1 \\ \Rc_2
\end{pmatrix}\,,\label{db_sys}
\end{equation}
where 
\begin{eqnarray}
&&\Wc_1=\frac{1+2n\ell d_i^2 f_1 k^2}{d_i^4 f_1^2}-\frac{1}{d_i^2} \,, \quad \Wc_2= -\frac{g_1}{d_i^2f_1}-\frac{1+2n\ell d_i^2 f_1 k^2}{d_i^4 f_1^2}\nn\\
&&\Wc_3=g_1f_1+\frac{1+2n\ell f_1 d_i^2 k^2}{d_i^2}\,, \quad \Wc_4=g_1^2-\frac{1-2n\ell g_1 d_i^2 k^2}{d_i^2}\,,\nn\\
&&\Rc_1=-\frac{f_0+g_0}{f_1 d_i^2}-\frac{m_1}{d_i^2f_1^2k^2}\,, \qquad \Rc_2=g_1 (f_0+g_0)+\frac{n_1}{k^2}+2n\ell k^2 (f_0+g_0)\,.
\end{eqnarray}
For $n\,,\ell \neq 0$ we can find a solution to this system assuming $m_1=n_1=f_0=g_0=0$ 
\begin{equation}
\varphi=\frac{\lambda_+ - g_1}{f_1} \psi_+ + \frac{\lambda_- - g_1}{f_1} \psi_- \,,\qquad \psi=\psi_++\psi_- \,, \label{db_flux}
\end{equation}
where $\psi_\pm$ are solutions of the equation
\begin{equation}
k^{-2}\Lc \psi_\pm=\lambda_\pm^2\psi_\pm+2n\ell \lambda_\pm k^2 \psi_\pm\,,
\label{helm_like}
\end{equation}
and parameters $\lambda_\pm$ are given by 
\begin{equation}
\lambda_\pm=\frac{1}{2}\left[\frac{1}{d_i^2 f_1} + g_1 \pm \sqrt{\left(\frac{1}{d_i^2 f_1} + g_1\right)^2 - 4 \frac{f_1 + g_1}{d_i^2 f_1}}\right]\,.
\end{equation}
Either solving \eqref{helm_like} directly or following the construction of \citet{Corsetti1973}  (see also \citet{CK1957}) we   obtain the following analytic solutions $\psi_\pm$:
\begin{eqnarray}
\psi_\pm&&=c_\pm\left[\ell J_0(\lambda_\pm r)-n r J_1(\lambda_\pm r)\right]\nn\\
&&+\sum_m a_m^\pm\left[\ell\lambda_\pm I_{\ell m}(\sigma_\pm r)+nr\frac{d}{dr}I_{\ell m}(\sigma_\pm r)\right]cos(mu)\,,
 \label{psi_beltrami}
\end{eqnarray}
where $\sigma_\pm:=\sqrt{m^2n^2-\lambda_\pm^2}$ and $I_{\ell m}$ denotes the modified Bessel function of the first kind with order $\ell m$. The parameters $c_\pm$ and $a_m^\pm$ can be specified in connection with the desirable boundary conditions. The functions $\psi_\pm$ are poloidal flux functions of helically symmetric Beltrami fields with Beltrami parameters $\lambda_\pm$. Their combination \eqref{db_flux} is a homogeneous solution of the system \eqref{db_sys}. Since the solution is a linear combination of two Beltrami fields, the resulting solution is called double-Beltrami (DB). Another reason for adopting this terminology is that the resulting velocity and magnetic fields satisfy conditions that involve the double curl operator. Such states, are not only natural solutions of the incompressible Hall MHD equilibrium equations (see \citet{Mahajan1998}) but they occur also as relaxed states via minimization principles \citep{Yoshida2002}. They have been used to construct high-beta equilibria with flows for 1D \citep{Mahajan1998,Iqbal2001} and axisymmetric systems \citep{Yoshida2001} but not for helically symmetric ones. Here we compute a helical DB equilibrium in view of \eqref{db_flux} and \eqref{psi_beltrami}. The computed configuration is depicted in figure \ref{fig_1}. The flux function $\psi$ labels the magnetic surfaces  while the function $\phi$ labels the ion flow surfaces. We obtained the configuration of Fig. \ref{fig_1}, possessing closed surfaces, for normalized ion skin depth $d_i=0.09$, $f_1=4.2$, $g_1=2.01$ and imposing the vanishing of $\psi$ on some predetermined boundary points, yielding the values of the free parameters in the truncated expansions \eqref{psi_beltrami}. We observe that the ion surfaces depart significantly from the electron-magnetic surfaces, though in a manner consistent with other computations for axisymmetric \citep{Guazzotto2015} and translationally symmetric \citep{Kaltsas2017} HMHD equilibria, resulting in a configuration with distinct helical structures for the ions and the electrons. 
\section{Conclusion}
\label{sec_VI}
We derived the helically symmetric extended magnetohydrodynamics Poisson bracket and the corresponding set of Casimirs which consists of four families of helically symmetric invariants. The Poisson bracket was employed in order to describe helical dynamics and the Casimirs with the Hamiltonian were used to derive, via an Energy-Casimir variational principle, the equilibrium  equations of helically symmetric XMHD. This  symmetry makes both the dynamical and equilibrium equations more involved than the corresponding translationally symmetric equations, through the presence of a scale factor $k$, and new purely helical contributions. The equilibrium equations were manipulated further for two simpler cases:  first was the axisymmetric barotropic and incompressible XMHD and second the helically symmetric barotropic and incompressible HMHD. Both systems with barotropic closure were cast in Grad-Shafranov-Bernoulli forms, which describe completely the respective equilibria.  In the incompressible cases the Bernoulli equation can no longer be derived via the standard EC principle but one has to return to the primary equations of the model. The Bernoulli equation decouples from the equilibrium PDE system,  becoming a secondary condition for the computation of the pressure. As an example,  a particular case of equilibria was studied by means of an analytical solution. The application concerns an incompressible, helically symmetric plasma described by HMHD,  for which we derived an analytic double-Beltrami solution and constructed an equilibrium configuration with non-planar helical axis which can be regarded as straight-stellarator-like equilibrium.

\section*{Acknowledgements}
This work has been carried out within the framework of the EUROfusion Consortium and has received funding from (a) the National Programme for the Controlled Thermonuclear Fusion, Hellenic Republic and (b) Euratom
research and training program 2014-2018 under Grant Agreement No.~633053. The views and opinions expressed
herein do not necessarily reflect those of the European Commission. DAK was supported by a Ph.D.\  grant from the Hellenic Foundation for Research and Innovation (HFRI) and the General Secretariat for Research and Technology (GSRT). DAK and GNT would like to thank George Poulipoulis, Apostolos Kuiroukidis and Achilleas Evangelias for very useful discussions on helically symmetric equilibria.  PJM  was supported by  the U.S. Department of Energy Contract DE-FG05-80ET-53088 and  a Forschungspreis from the Alexander von Humboldt Foundation. He  warmly acknowledges the hospitality of the Numerical Plasma Physics Division of  Max Planck IPP,  Garching where a portion of this research was done. 

\numberwithin{equation}{section}


\appendix
\section{Compressible helically symmetric XMHD dynamics}
\label{app_A}
The compressible XMHD dynamics that respect helical symmetry is given by $\partial_t \eta_{_{HS}}=\{\eta_{_{HS}},\Hc\}^{^{XMHD}}_{_{HS}}$.
In view of \eqref{hs_poisson} and \eqref{hs_hamiltonian} we have:
\begin{eqnarray}
\partial_t\rho&&=-\nb\bcdot(\rho\nb\Upsilon)+[\chi,\rho]\,,\label{app_1}\\
\partial_t v_h&&=\rho^{-1}k\left([\Hc_\Omega,k^{-1}v_h]+[k^{-1}B_h,\psi^*]+\nb(k^{-1}v_h)\bcdot\nb\Hc_w\right)\,, \label{app_2}\\
\partial_t\Omega&&=[\Hc_\Omega,\rho^{-1}\Omega]-2n\ell [\Hc_\Omega,\rho^{-1}k^3v_h]+\nb\bcdot(\rho^{-1}\Omega\nb \Hc_w)-2n\ell \nb\bcdot(\rho^{-1}k^3v_h\nb \Hc_w) \nn\\
&&+[kv_h,k^{-1}v_h]+[k^{-1}B_h,\rho^{-1}k B_h^*]+[\rho^{-1}\Lc\psi,\psi^*]-2n\ell[\rho^{-1}k^3B_h,\psi^*]\,, \label{app_3}\\
\partial_t w&&=-\Delta\Hc_\rho+[\Hc_w,\rho^{-1}k^{-2}\Omega]-2n\ell [\Hc_w,\rho^{-1}k v_h]-\nb\bcdot(\rho^{-1}\Omega\nb \Hc_\Omega) \nn \\
&&+2n \ell \nb \bcdot(\rho^{-1}k^3v_h\nb \Hc_\Omega) +\nb\bcdot\left(kv_h\nb(k^{-1}v_h)\right)-\nb\bcdot(\rho^{-1}k B_h^*\nb(k^{-1}B_h))\nn \\
&&+\nb\bcdot(\rho^{-1}\Lc\psi\nb\psi^*)-2n\ell\nb\bcdot(\rho^{-1}k^3B_h\nb\psi^*)\,, \label{app_4}\\
\partial_t B_h^*&&= k^{-1}\big([\Hc_\Omega,\rho^{-1}kB_h^*]+\nb\bcdot(\rho^{-1}kB_h^*\nb\Hc_w)+[kv_h,\psi^*]\nn \\
&&-2n\ell\rho^{-1}k^4[\Hc_\Omega,\psi^*]-2n\ell \rho^{-1}k^4\nb\psi^*\bcdot\nb\Hc_w +d_i[\rho^{-1}kB_h^*,k^{-1}B_h]\nn \\
&&-d_i[\rho^{-1}\Lc\psi,\psi^*]-2n\ell d_i\rho^{-1}k^4[\psi^*,k^{-1}B_h]+2n\ell d_i[\rho^{-1}k^3B_h,\psi^*] \nn\\
&& +d_e^2[k^{-1}B_h,\rho^{-1}\Omega]-2n\ell d_e^2[k^{-1}B_h,\rho^{-1}k^3v_h]+d_e^2[\rho^{-1}\Lc\psi,k^{-1}v_h] \nn \\
&&-2n\ell d_e^2\rho^{-1}k^4[k^{-1}B_h,k^{-1}v_h]-2n\ell d_e^2[\rho^{-1}k^3B_h,k^{-1}v_h]\big)
\,,\label{app_5}\\
\partial_t\psi^*&=&\rho^{-1}\left([\Hc_\Omega,\psi^*]+\nb\psi^*\bcdot\nb\Hc_w+d_i[\psi^*,k^{-1}B_h]+d_e^2[k^{-1}B_h,k^{-1}v_h]\right)\,,\label{app_6} 
\end{eqnarray}
where $\Hc_\rho$ is given by \eqref{func_der_3} while $\Hc_\Omega$ and $\Hc_w$ are given by \eqref{func_der_6}. For incompressible plasma ($\rho=1$, $w=0$) the terms that contain functional derivatives $F_\rho$ and $F_w$ in \eqref{hs_poisson} cease to exist. Hence \eqref{app_1} and \eqref{app_4} are absent, while $\Hc_w=0$ and $\Hc_\Omega=\chi$ in the rest of the equations, leading to the system \eqref{dyn_vh}-\eqref{dyn_psi} for the incompressible dynamics. 
\medskip


\bibliographystyle{jpp}

\bibliography{biblio}

\end{document}